\documentclass[twocolumn]{aastex631}

\usepackage{graphicx}
\usepackage{amsmath}
\usepackage{cleveref}
\usepackage{float}
\usepackage{pgffor}
\usepackage{tabularx}
\usepackage{array}
\usepackage{xcolor}
\newcolumntype{Y}{>{\raggedright\arraybackslash}X}

\usepackage{makecell}

\begin{document}

\title{A Two-level Radial-velocity Zero-point Calibration for LAMOST MRS with Gaia and APOGEE and a Value-added RV Catalogue}

\shorttitle{Two-level RVZP Calibration and Value-added RV Catalogue for LAMOST MRS}
\shortauthors{Zhang \& Yuan}

\author[0009-0005-7743-6229]{Jinming Zhang}
\affiliation{Institute for Frontiers in Astronomy and Astrophysics, Beijing Normal University, Beijing, 102206, China}
\affiliation{School of Physics and Astronomy, Beijing Normal University, Beijing, 100875, China}

\author[0000-0003-2471-2363]{Haibo Yuan}
\affiliation{Institute for Frontiers in Astronomy and Astrophysics, Beijing Normal University, Beijing, 102206, China}
\affiliation{School of Physics and Astronomy, Beijing Normal University, Beijing, 100875, China}

\correspondingauthor{Haibo Yuan}
\email{yuanhb@bnu.edu.cn}

\begin{abstract}
The LAMOST Medium-Resolution Survey (MRS) provides a large stellar spectroscopic data set for Galactic kinematics and time-domain radial-velocity (RV) studies. However, the multi-spectrograph, multi-exposure, multi-fiber observing strategy can imprint RV zero-point (RVZP) systematics that vary across instrumental and temporal hierarchies. 
We construct a two-level empirical RVZP correction to LAMOST DR12 MRS using Gaia DR3 magnitude--colour-corrected RVs and APOGEE DR17 RVs as external references: a spectrograph--exposure correction in \texttt{(lmjm, planid, spid)} units, followed by a fiber--time correction in \texttt{(spid, fiberid, time\_tag)} units. Within each unit, RVZPs are estimated from Gaia and APOGEE residuals using a weighted-median estimator, and are subtracted from the pipeline RVs to obtain corrected velocities. 
For high-S/N spectra, the scatter relative to APOGEE DR17 decreases from $\sim$1.1\,km\,s$^{-1}$ to $\sim$0.52\,km\,s$^{-1}$ after the full two-level correction, implying a $\sim$2\,$\times$ improvement in RV precision.
We release a value-added catalogue of two-level-corrected RVs for 11,129,477 blue-arm spectra, including 8,158,271 single-exposure spectra and 2,971,206 coadded spectra, enabling consistent RV analyses across exposures, spectrographs, fibers, and time.
\end{abstract}

\section{Introduction}

Radial velocity (RV) is the line-of-sight component of stellar motion. Large RV samples enable cluster membership determination, the identification of binaries and pulsators, black-hole candidate searches, and multi-epoch time-domain spectroscopy. These applications require RV catalogues with not only small random errors, but also well-controlled zero points across observational hierarchies.

The Large Sky Area Multi-Object Fiber Spectroscopic Telescope (LAMOST; \citealt{cui2012lamost,deng2012lamost,zhao2012lamost,Liu_2014}) is a wide-field, multi-object fiber spectroscopic facility that collects spectra simultaneously with multiple spectrographs and thousands of fibers. The LAMOST medium-resolution survey (MRS; \citealt{lamostMRS}) has a typical resolving power of $R \sim 7500$ and delivers blue- and red-arm spectra. Its emphasis on multi-epoch RV science and time-domain kinematics makes it particularly sensitive to RV zero-point differences among exposures, spectrographs, and fibers.

RV errors arise not only from template matching or cross-correlation, but also from calibration and instrumental effects. Wavelength calibration maps CCD pixels to physical wavelengths, and small errors in the wavelength solution propagate directly into line-position shifts and RV offsets. Additional contributors include non-simultaneity between arc-lamp and science exposures, thermo-mechanical variations, spectrograph drift, fiber-dependent effects, residual background/sky subtraction, and reference-frame differences between surveys. Previous MRS analyses have shown that spectrograph exposure units (SEUs) exhibit time-dependent RVZPs. Gaia DR3 RVS validation likewise demonstrates the importance of accounting for magnitude- and colour-dependent zero points and other systematics (e.g., straylight, PSF/LSF evolution, and crowding; \citealt{katz2023,blomme2023}). We therefore treat the RV zero point as an empirical correction term rather than attempting to attribute it to a single physical mechanism.

\citet{wangR2019MRS} established an early baseline for LAMOST MRS RVs using template matching, independent blue- and red-arm measurements, and RV standard stars to estimate static zero points, demonstrating km\,s$^{-1}$-level precision for million-scale spectra. \citet{zhangBO2021MRS} further showed that static zero points are inadequate for multi-epoch time-domain studies and constructed an SEU/exposure-level time-dependent RVZP correction using Gaia references and repeat observations.

Recently, \citet{zhang2026lrs} developed a hierarchical RV calibration scheme for LAMOST low-resolution spectra (LRS) that separates RV systematics into two components: (i) internal wavelength-solution inconsistencies within spectra and (ii) global zero-point offsets. These effects are corrected in distinct layers rather than being absorbed into a single global term. For LRS, the procedure first corrects broad-band, wavelength-dependent inconsistencies and then calibrates the remaining zero-point systematics across the instrumental and temporal hierarchies, yielding substantially improved internal RV consistency.

Here we follow this philosophy and extend it to LAMOST medium-resolution spectra (MRS), with one key simplification. Because the MRS wavelength coverage is much narrower than in LRS, we skip the broad-band wavelength-consistency step and apply only the zero-point calibration layer, using Gaia DR3 and APOGEE DR17 RVs as external references.

The paper is organized as follows. Section~2 describes the LAMOST MRS data and the Gaia and APOGEE reference data. Section~3 presents the two-level RVZP methodology. Section~4 shows the resulting RVZP structures at the spectrograph--exposure and fiber--time levels. Section~5 validates the precision improvement using repeat observations, reference comparisons, and an independent DR19 test for the single-exposure spectra, and presents  uncertainty estimation for the corrected RVs. Section~6 documents the data-product columns and recommended usage, Section~7 summarizes the conclusions, and Appendix~A presents the corresponding calibration for the coadded spectra.

\section{Data}

\subsection{LAMOST DR12 v1.1 MRS}

This work uses the LAMOST DR12 v1.1 medium-resolution spectroscopic (MRS) dataset, spanning September 2017 to June 2024. Each observation provides blue- and red-arm spectra covering [4950\,\AA, 5350\,\AA] and [6300\,\AA, 6800\,\AA], respectively, at $R \sim 7500$.

In DR12 v1.1 MRS, the LAMOST Stellar Parameter Pipeline (LASP; \citealt{wu2011LASP}) reports two RVs: the directly measured $RV_{\rm LASP0}$ and $RV_{\rm LASP1}$ after a static spectrograph correction. We use $RV_{\rm LASP0}$ as input to our calibration and treat $RV_{\rm LASP1}$ as the pipeline-corrected RV. We define
\[ \mathrm{RVZP} = \mathrm{RV}_{\rm LASP0} - \mathrm{RV}_{\rm reference}. \]

We calibrate only the blue arm because red-arm LASP RVs are not sufficiently reliable for this analysis. The application sample therefore consists of spectra that (1) are MRS blue-arm spectra and (2) have valid LASP RV measurements; it includes both single-exposure and coadded spectra.

We separate the calibration and application samples: the calibration sample is a high-reliability set of stellar spectra used to infer the RV zero points (excluding likely variables, binaries, and other problematic/non-stellar targets), while the inferred correction is applied to the broader application sample. For the calibration sample, we do not apply a hard cut on quoted RV uncertainties; instead, we use inverse-variance weighting to down-weight noisier spectra while retaining a larger reference sample.

Single-exposure and coadded spectra are processed with the same two-level RVZP framework (same reference system, weighting, and clipping) but are calibrated independently, reflecting that coadds combine consecutive same-night plate exposures whereas single-exposure spectra preserve exposure-level information. We show single-exposure results in the main text and the coadded calibration in Appendix~A.

In the data-product recommendation, \textbf{\texttt{rv\_lasp2}} denotes the final RV after applying all corrections.

\subsection{Gaia DR3 magnitude--colour-corrected RV reference}

Gaia is an ESA astrometry mission providing precise positions, proper motions, parallaxes, and radial velocities \citep{gaia2016gaia}. Its Radial Velocity Spectrometer (RVS) delivers the Gaia DR3 RV catalogue from the first 34 months of scanning spectra \citep{cropper2018gaia,gaiadr3}. RVS covers 847--874\,nm; Gaia DR3 provides combined RVs for 33,812,183 stars, with a characteristic precision of $\sim$1.3\,km\,s$^{-1}$ for bright sources. Gaia's all-sky, uniform measurements make it a convenient external RV reference for large-scale calibration.

Because Gaia DR3 RV zero points depend on magnitude and colour \citep{katz2023,blomme2023}, we adopt the corresponding corrected Gaia RVs as our primary reference; these corrections are tied to the APOGEE system, ensuring consistency with the APOGEE DR17 reference.

\subsection{APOGEE DR17 reference}

The Apache Point Observatory Galactic Evolution Experiment (APOGEE; \citealt{majewski2017apogee}) is a high-resolution near-infrared SDSS survey using a 2.5\,m telescope and an H-band spectrograph ($R\sim22{,}500$). APOGEE DR17 released spectra for $\sim$650,000 stars with typical RV precision better than 0.1\,km\,s$^{-1}$. We use APOGEE DR17 as a second external reference; 1,093,347 LAMOST MRS spectra have APOGEE counterparts. APOGEE lacks all-sky coverage but complements Gaia with much higher RV precision.

If an MRS spectrum matches both Gaia and APOGEE, we include residuals to each reference within the same calibration unit, allowing the weighted-median solution to combine the two constraints.

\section{Method}

\subsection{Overview of calibration}

\begin{figure*}[htbp]
\centering
\makebox[\textwidth][c]{\includegraphics[width=0.9\textwidth]{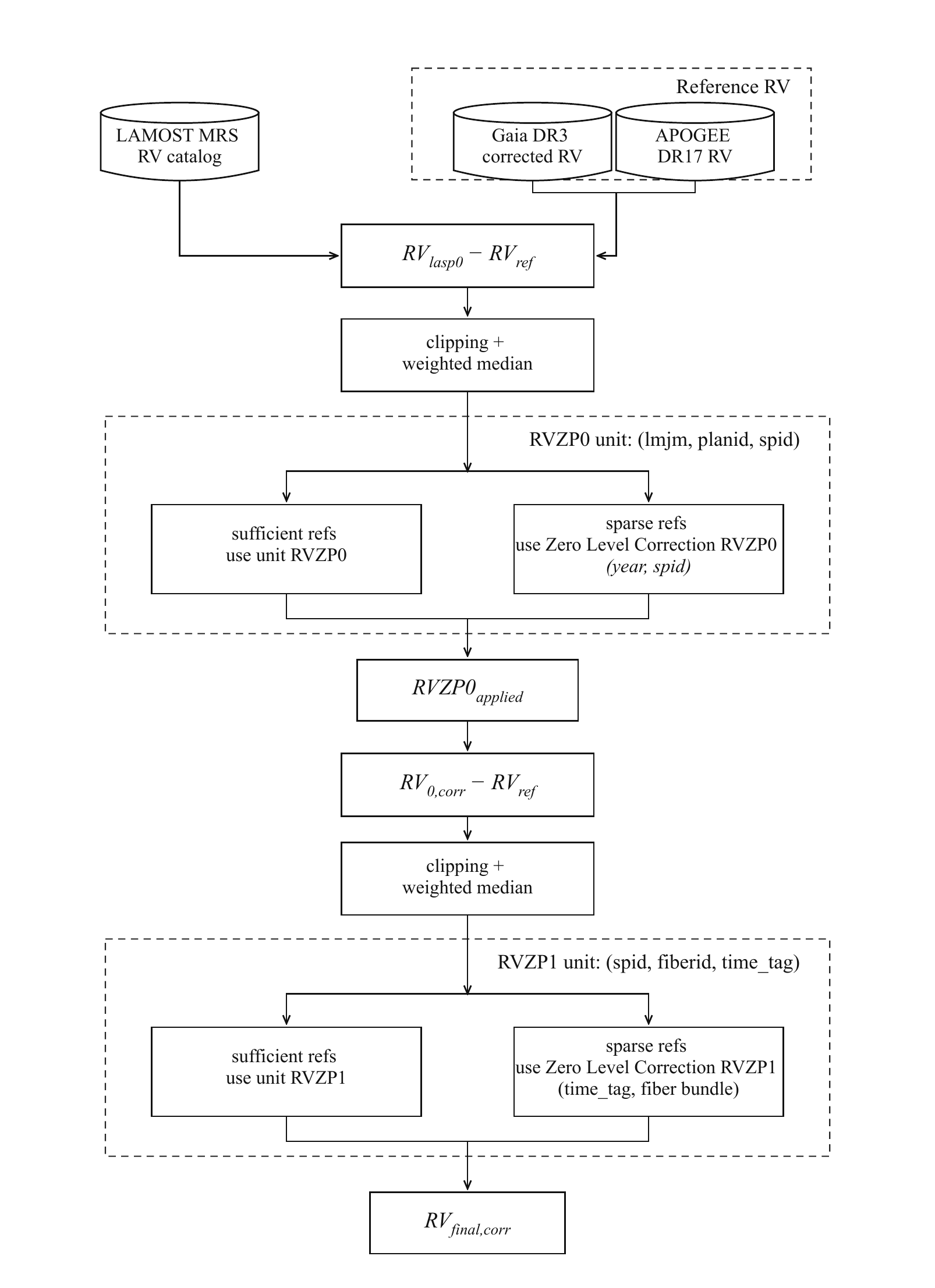}}
\caption{Flowchart of the two-level RV zero-point correction for LAMOST MRS. We compare LAMOST MRS RVs with Gaia-corrected and APOGEE DR17 RVs to derive (1) a spectrograph--exposure correction in \texttt{(lmjm, planid, spid)} units and (2) a fiber--time correction inferred from the good sample after applying the first-level correction.}
\label{fig:figure_1}
\end{figure*}

We adopt a hierarchical zero-point calibration. At the first level, we estimate a spectrograph--exposure correction that captures the empirical RV zero point shared within a given exposure (plate and spectrograph). At the second level, using the good sample after the first-level correction, we estimate a fiber--time correction to remove residual systematics associated with fiber and observing time. The final output is the two-level-corrected velocity \textbf{$RV_{\rm corr}^{\rm sp,fib}$}. This hierarchy is motivated by the observing structure of LAMOST MRS: RV systematics can depend on exposure, spectrograph, fiber position, and observing season, and thus cannot be represented by a single global constant. Figure~\ref{fig:figure_1} summarizes this workflow.

\subsection{Spectrograph-exposure zero point}

We define a spectrograph--exposure correction unit as \texttt{(lmjm, planid, spid)}, i.e., a unique combination of exposure, plate, and spectrograph. This choice follows the observing structure of LAMOST MRS: effects such as wavelength-calibration residuals, short-term instrumental drift, and observing conditions are expected to be shared within the same spectrograph during a given exposure. The corresponding correction is therefore an empirical RV zero point at the spectrograph--exposure level.

Within each unit, we form external RV residuals in the common reference frame using Gaia-corrected RVs and APOGEE DR17 RVs:

\[ \Delta RV_{\rm Gaia} = RV_{\rm LASP0} - RV_{\rm Gaia}, \]

\[ \Delta RV_{\rm APOGEE} = RV_{\rm LASP0} - RV_{\rm APOGEE\,DR17}. \]

These residuals follow our convention $\mathrm{RVZP}=\mathrm{RV}_{\rm LASP0}-\mathrm{RV}_{\rm reference}$; the corrected MRS velocity is obtained by subtracting the estimated zero point from the measured RV.

We assign each residual an inverse-variance weight,

\[ w = \frac{1}{\sigma_{\rm LASP}^2+\sigma_{\rm ref}^2}, \]

with

\[ \sigma_{\rm LASP}=\max\left(\sigma_{RV_{\rm LASP0}},\ 0.5\ {\rm km\ s^{-1}}\right). \]

The 0.5\,km\,s$^{-1}$ error floor prevents unrealistically small formal uncertainties from dominating the unit-level estimate.

For each unit we compute an initial median and MAD, clip outlying residuals to reduce contamination from RV-variable sources, and then compute a weighted median from the retained sample to obtain the spectrograph--exposure zero point.

We do not deduplicate the Gaia and APOGEE reference samples. If a calibration star has both a Gaia-corrected RV and an APOGEE RV, it contributes one residual from each catalogue to its calibration unit, with each residual weighted by its propagated uncertainty. Because the two reference sets have been calibrated onto a common velocity system, the impact of this overlap on the inferred zero point is small. First, APOGEE RVs are substantially more precise, most of the zero-point information from an overlapping star is carried by the more highly weighted APOGEE residual. Second, overlapping stars typically constitute only $\sim17\%$ of the reference sample in a given unit and thus have little influence on the weighted-median zero point. 

We exclude units with fewer than 20 contributions. There are 136,320 spectrograph--exposure units in total, of which 117,428 satisfy this requirement.

For excluded units we adopt a ``zero-level'' correction to stabilize low-evidence cases. Specifically, we take the weighted median of well-sampled spectrograph--exposure corrections within broader groups and apply that value to the excluded units. These broader groups are defined by observing year and spectrograph, as motivated by the empirical structure shown in Figure~\ref{fig:figure_2} and discussed in Section~4.1. This procedure mitigates over-reaction to small-number statistics and affects about 2.6\% of the spectra, typically due to a lack of effective reference matches (e.g., weather or very bright time).

\subsection{Fiber-time zero point}

The fiber--time correction unit is defined as \texttt{(spid, fiberid, time\_tag)}, i.e., a unique combination of spectrograph, fiber position, and observing-time tag. This second level targets residual systematics that can persist after the spectrograph--exposure correction, capturing spatial--temporal structure across fibers and time segments within a given spectrograph rather than an additional exposure-wide offset.

We estimate the fiber--time correction only from spectra belonging to spectrograph--exposure units with sufficient sample size, to avoid propagating poorly constrained first-level zero points into the second-level correction.

The reference residuals, weights, clipping, and weighted-median estimator are the same as in the first level, but are computed using the first-level-corrected velocities: \textbf{$RV_{\rm corr}^{\rm sp}-RV_{\rm Gaia}$} and \textbf{$RV_{\rm corr}^{\rm sp}-RV_{\rm APOGEE\,DR17}$}.

\section{Results: zero-point structures}

\subsection{Structure of the spectrograph-exposure-level correction}

\begin{figure*}[htbp]
\centering
\includegraphics[width=0.95\textwidth]{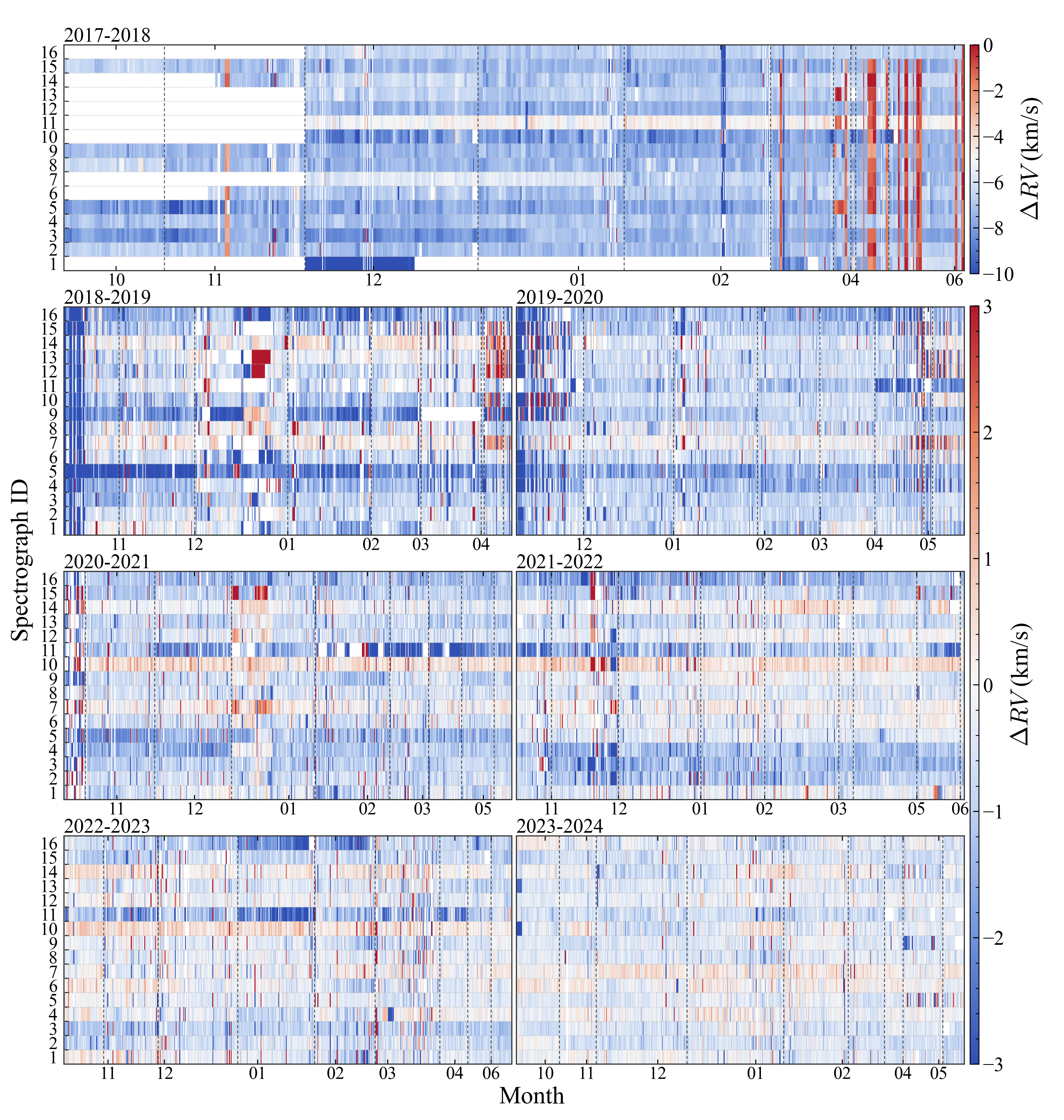}
\caption{Seasonal heatmap of the spectrograph--exposure RV zero-point correction, shown by observing season. In each panel, rows correspond to spectrographs, the x-axis is time, and each cell is one spectrograph--exposure unit (one LMJM exposure). Colors show the correction (km\,s$^{-1}$). We use \texttt{[-10,0]} for 2017--2018 because its zero-point level is offset relative to later seasons; all other seasons use \texttt{[-3,3]}. Gray vertical dashed lines mark the starts of MRS observing cycles within the alternating MRS/LRS schedule. The plot illustrates the strong time- and spectrograph-dependent structure that motivates the spectrograph--exposure correction.}
\label{fig:figure_2}
\end{figure*}

Figure~\ref{fig:figure_2} demonstrates that the spectrograph--exposure RVZP is not a single global constant: it shows strong, coherent structure with observing season, date, and spectrograph, with amplitudes at the km\,s$^{-1}$ level. These patterns can bias high-S/N RV comparisons across dates or spectrographs, motivating an explicit spectrograph--exposure correction as the first (dominant) level before solving for smaller fiber--time residuals.

The gray dashed lines mark the starts of MRS observing blocks within the alternating MRS/LRS schedule. MRS blocks typically last \ensuremath{\sim}half a month and are separated by LRS blocks during which the spectrographs are switched. For several spectrographs, RVZP patterns change across these boundaries, consistent with partial resets of the instrumental state; for others, the month-scale pattern remains similar across successive blocks.

We treat the spectrograph--exposure correction as an empirical RV zero point that absorbs a mixture of effects tied to the exposure and instrument state (e.g., observing conditions and wavelength-calibration residuals). The heatmap constrains the structure but not its physical origin; attributing the patterns to specific causes (lamp state, environmental drift, maintenance events, pipeline factors) would require additional calibration information beyond the scope of this work.

Figure~\ref{fig:figure_2} also motivates the grouping used for the ``zero-level'' correction of low-evidence spectrograph--exposure cells. Because different spectrographs follow distinct RVZP patterns while each spectrograph retains year-scale continuity, we group by observing year and spectrograph. A month-by-month grouping is not adopted because some MRS blocks are too sparse to define stable monthly levels for all spectrographs and would require ad hoc merge/split choices; year\,$\times$\,spectrograph is a more stable and reproducible compromise.

\subsection{Structure of the fiber--time correction}

We estimate the second-level correction in \texttt{(spid, fiberid, time\_tag)} units to capture residual systematics tied to fiber position and slowly varying instrument state within each spectrograph. The time axis is discretized via \texttt{time\_part}, which splits each observing year into \texttt{P1 = Sep--Nov}, \texttt{P2 = Dec--Feb}, and \texttt{P3 = Mar--Jun}; we use \texttt{time\_tag} to denote the resulting time bin (observing-year label plus \texttt{time\_part}). An observing year runs from September to June, with Jan--Jun assigned to the previous year (e.g., \texttt{2017 P2} corresponds to early 2018). This segmentation balances time resolution against reference statistics.

\begin{figure*}[htbp]
\centering
\includegraphics[width=0.95\textwidth]{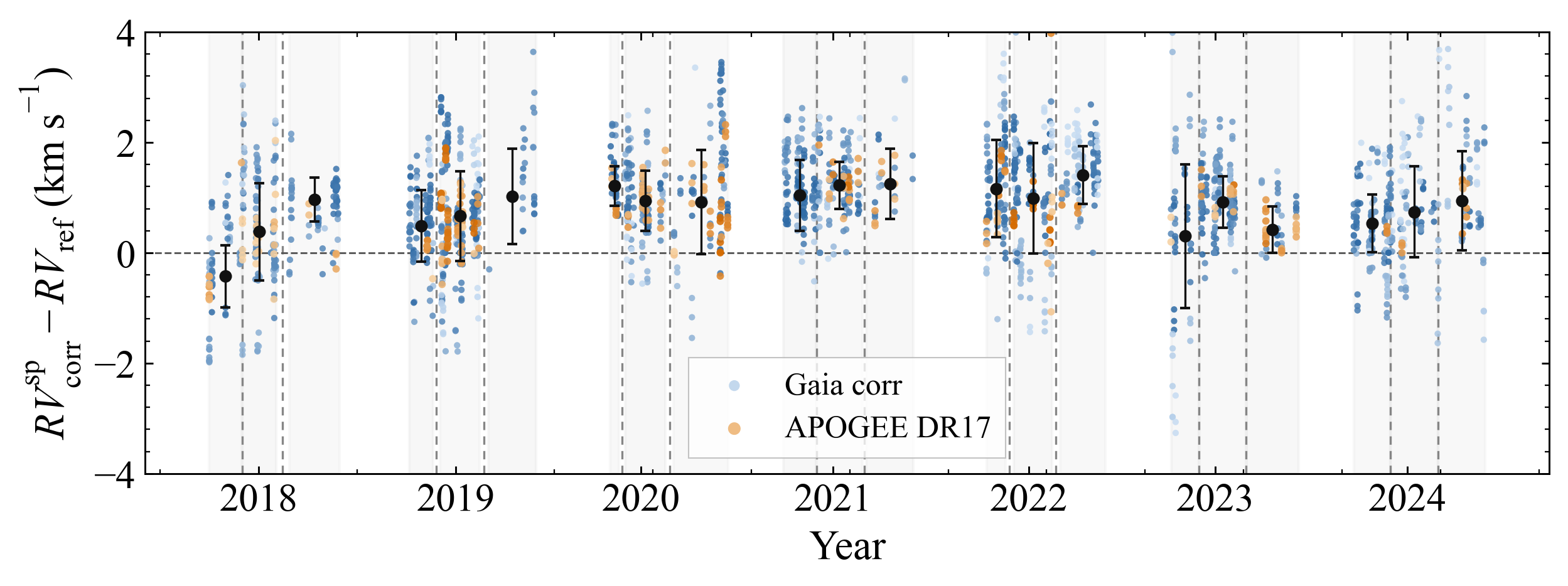}
\caption{Residuals versus LMJM for the fiber with the most external-reference matches. The y-axis shows $RV_{\mathrm{corr}}^{\mathrm{sp}}-RV_{\mathrm{ref}}$, i.e., the MRS RV after the spectrograph--exposure correction relative to the external reference. Blue points use Gaia-corrected RVs and orange points use APOGEE DR17 RVs; point darkness encodes the weighted-median weight (darker = higher weight). Black filled circles show the weighted-median residual for each fiber--time unit with its $\sigma$ error bar. Known and candidate double-lined spectroscopic binaries (SB2) from \citet{kovalev2024} are excluded. The remaining coherent time dependence motivates the second-level fiber--time correction.}
\label{fig:figure_3}
\end{figure*}

Figure~\ref{fig:figure_3} illustrates that, even after the spectrograph--exposure correction, coherent time-dependent residuals can remain within a single fiber. We define \textbf{84,000} fiber--time units in total; units with fewer than 17 external-reference contributions are excluded from the direct estimate, leaving 55,451 units in the main calculation.

\begin{figure*}[htbp]
\centering
\includegraphics[width=0.95\textwidth]{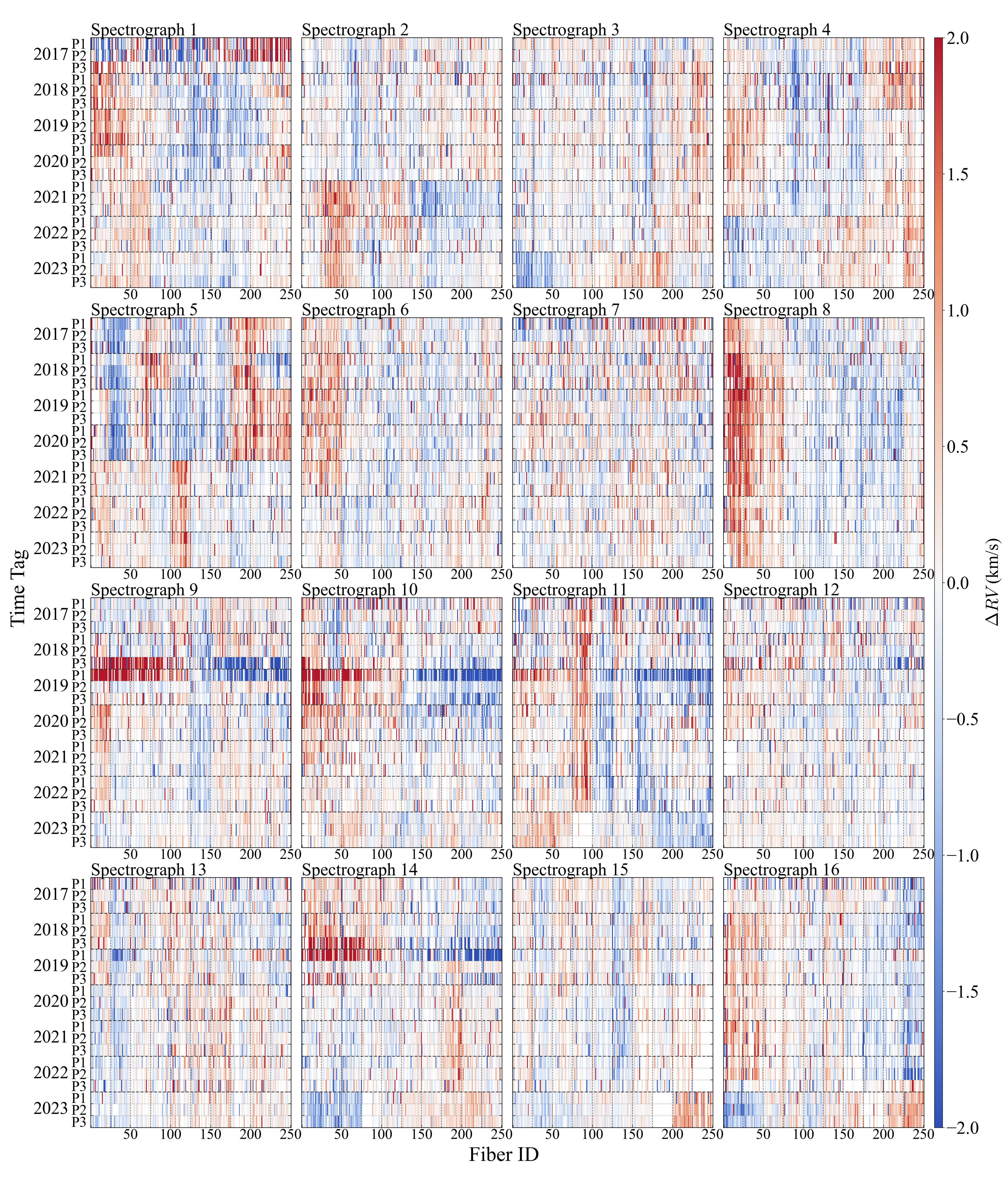}
\caption{Fiber--time RV zero-point correction map. Each panel shows one spectrograph: the x-axis is \texttt{fiberid} (250 fibers) and the y-axis orders \texttt{time\_tag} bins from early to late, with \texttt{P1--P3} marking the three segments of each observing year (Sep--Jun; e.g., \texttt{2017 P2} falls in early 2018). Each cell is one \texttt{(spid, fiberid, time\_tag)} correction for the corresponding spectrograph panel, color-coded in km\,s$^{-1}$ over \texttt{[-2,2]}. Gray vertical dashed lines separate 25-fiber bundles and gray horizontal dashed lines separate observing years. The remaining coherent structure after the spectrograph--exposure correction motivates the second-level fiber--time correction.}
\label{fig:figure_4}
\end{figure*}

Figure~\ref{fig:figure_4} shows that structured residuals persist as a function of \texttt{fiberid} and \texttt{time\_tag} within each spectrograph. Stripes and localized patches indicate that the remaining RVZP is not dominated by uncorrelated noise; patterns vary among spectrographs, and features can persist across multiple \texttt{time\_tag} bins or be confined to a single segment.
The map shows sharp discontinuities at some bundle boundaries, indicating that bundle-scale systematics contribute substantially to the residual structure.

For low-evidence units (\texttt{n < 17}), we adopt a grouped baseline (``zero-level'') correction inferred from well-sampled units.
Monthly bins are physically appealing because the MRS/LRS alternating schedule runs in $\sim$month-long blocks and Figure~\ref{fig:figure_3} indicates that residual variations occur on timescales finer than a single \texttt{time\_part} segment. However, a month-by-month split combined with fiber-level subdivision leaves too many cells sparsely populated to provide a stable baseline.
We therefore group by \texttt{time\_part} (three segments per observing year) and 25-fiber bundles, which captures the dominant large-scale structure while retaining sufficient reference statistics for robust weighted-median estimates. Because fibers within a bundle are not strictly identical, this grouped baseline is used only as a coarse fallback for sparse units, not as a substitute for the full \texttt{(fiberid, time\_tag)} correction. It is applied to 2.14\% of spectra.


\section{Verification: precision and external checks}

\subsection{Precision from repeat observations and training references}

\begin{figure*}[htbp]
\centering
\includegraphics[width=0.95\textwidth]{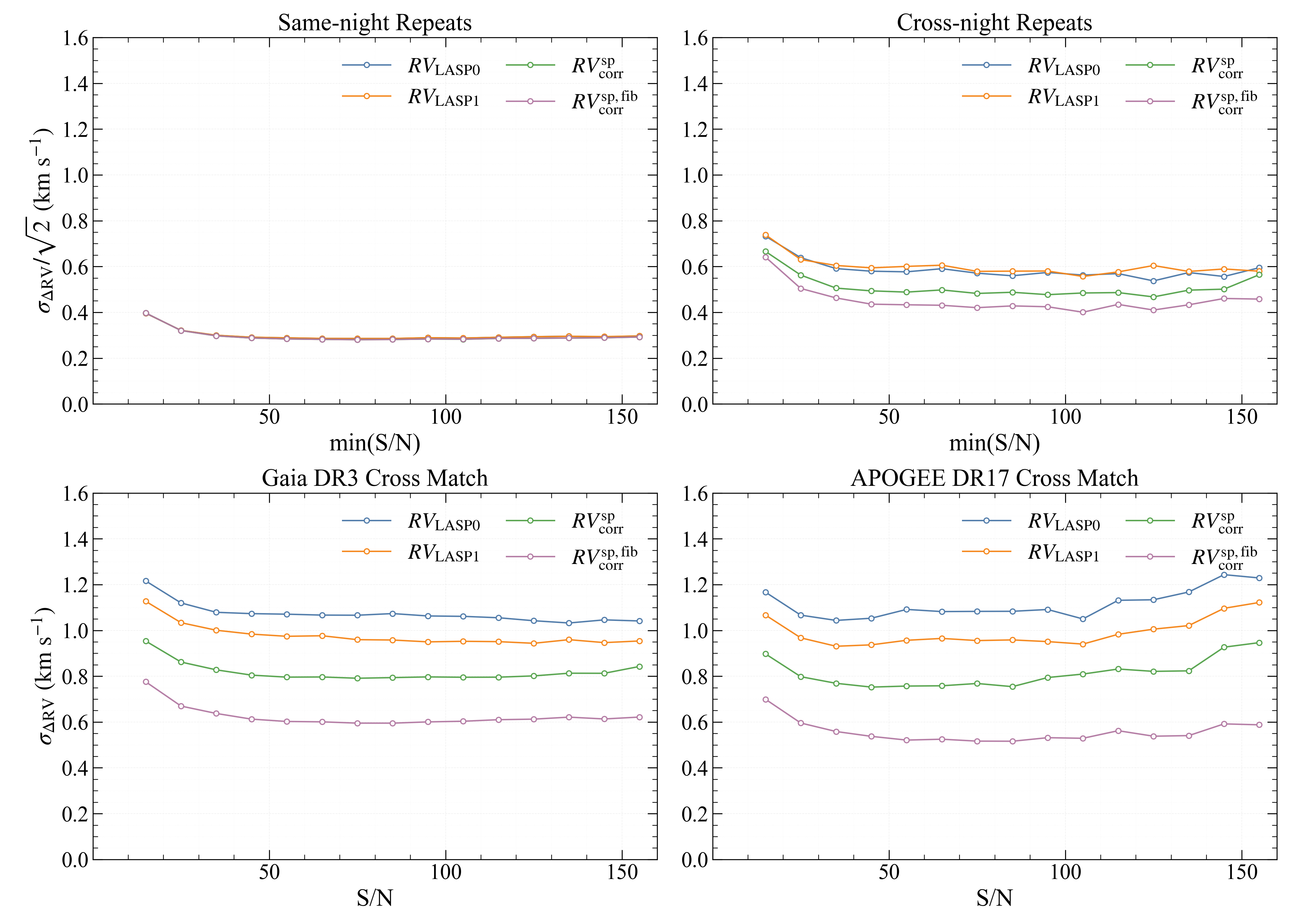}
\caption{Residual scatter versus LAMOST MRS S/N in four validation tests. From left to right and top to bottom, panels show same-night repeats, cross-night repeats, residuals relative to the Gaia DR3 corrected RVs, and residuals relative to APOGEE DR17. Curves show $RV_{\rm LASP0}$ (blue), $RV_{\rm LASP1}$ (orange), $RV_{\rm corr}^{\rm sp}$ (green), and $RV_{\rm corr}^{\rm sp,fib}$ (purple). In each S/N bin, we estimate $\sigma$ by fitting a Gaussian to the core of the residual distribution, applying iterative 3$\sigma$ clipping until convergence. For repeat observations, $\sigma$ is measured from pairwise RV differences for the same source (binned by the lower S/N) and divided by $\sqrt{2}$ to report the single-observation RV precision. For Gaia/APOGEE, $\sigma$ is computed from LAMOST--reference RV differences in S/N bins of width 10 (for Gaia, $\texttt{rv\_gaiadr3\_err}<0.5$\,km\,s$^{-1}$) and is not divided because it includes only small external reference uncertainties. The two-level correction improves both repeatability and external consistency, with the largest gains in cross-night and external-reference tests.}
\label{fig:figure_5}
\end{figure*}

Figure~\ref{fig:figure_5} shows that the two-level correction improves both internal repeatability and external consistency, with the largest gains in the cross-night and external-reference tests.

For same-night repeats, all four RV estimates converge to a single-observation precision of $\sim$0.28\,km\,s$^{-1}$ in the \texttt{S/N = 50 -- 60} bin, which we take as an internal precision floor under nearly identical conditions. For cross-night repeats in the same S/N bin, the single-observation precision improves from 0.60\,km\,s$^{-1}$ ($RV_{\rm LASP0}$) and 0.61\,km\,s$^{-1}$ ($RV_{\rm LASP1}$) to 0.50\,km\,s$^{-1}$ after the spectrograph--exposure correction and to 0.48\,km\,s$^{-1}$ after applying both levels.

External comparisons show similar improvement. In the \texttt{S/N = 50 -- 60} bin, the scatter relative to the Gaia magnitude--colour-corrected RVs (requiring $\texttt{rv\_gaiadr3\_err}<0.5$\,km\,s$^{-1}$) decreases from 1.07\,km\,s$^{-1}$ ($RV_{\rm LASP0}-RV_{\rm Gaia}$) to 0.80\,km\,s$^{-1}$ after the first-level correction and to 0.60\,km\,s$^{-1}$ after both levels. Relative to APOGEE DR17, the scatter decreases from 1.09\,km\,s$^{-1}$ to 0.76\,km\,s$^{-1}$ and to 0.52\,km\,s$^{-1}$. 
Thus, in this representative high-S/N APOGEE DR17 comparison, the final corrected RV scatter is roughly halved relative to the original LASP0 value, implying a $\sim$2\,$\times$ improvement in RV precision.
The slightly larger Gaia-based scatter at high S/N is expected given the lower RV precision of Gaia DR3 RVS compared with APOGEE.

At high S/N, the cross-night single-observation precision of $RV_{\rm corr}^{\rm sp,\,fib}$ ($\sim$0.48\,km\,s$^{-1}$) agrees with the APOGEE-based external scatter ($\sim$0.52\,km\,s$^{-1}$), once the APOGEE RV uncertainties ($\sim$0.10\,km\,s$^{-1}$) are taken into account.

A residual gap persists between the cross-night single-observation precision ($\sim$0.48\,km\,s$^{-1}$) and the same-night precision floor ($\sim$0.28\,km\,s$^{-1}$). This $\sim$0.20\,km\,s$^{-1}$ offset points to intermediate-timescale systematics not resolved by our correction scheme, plausibly arising from wavelength-dependent residuals and the coarse temporal sampling of the fiber--time correction (three 3--4 month segments per observing year).

\begin{figure*}[htbp]
\centering
\includegraphics[width=0.95\textwidth]{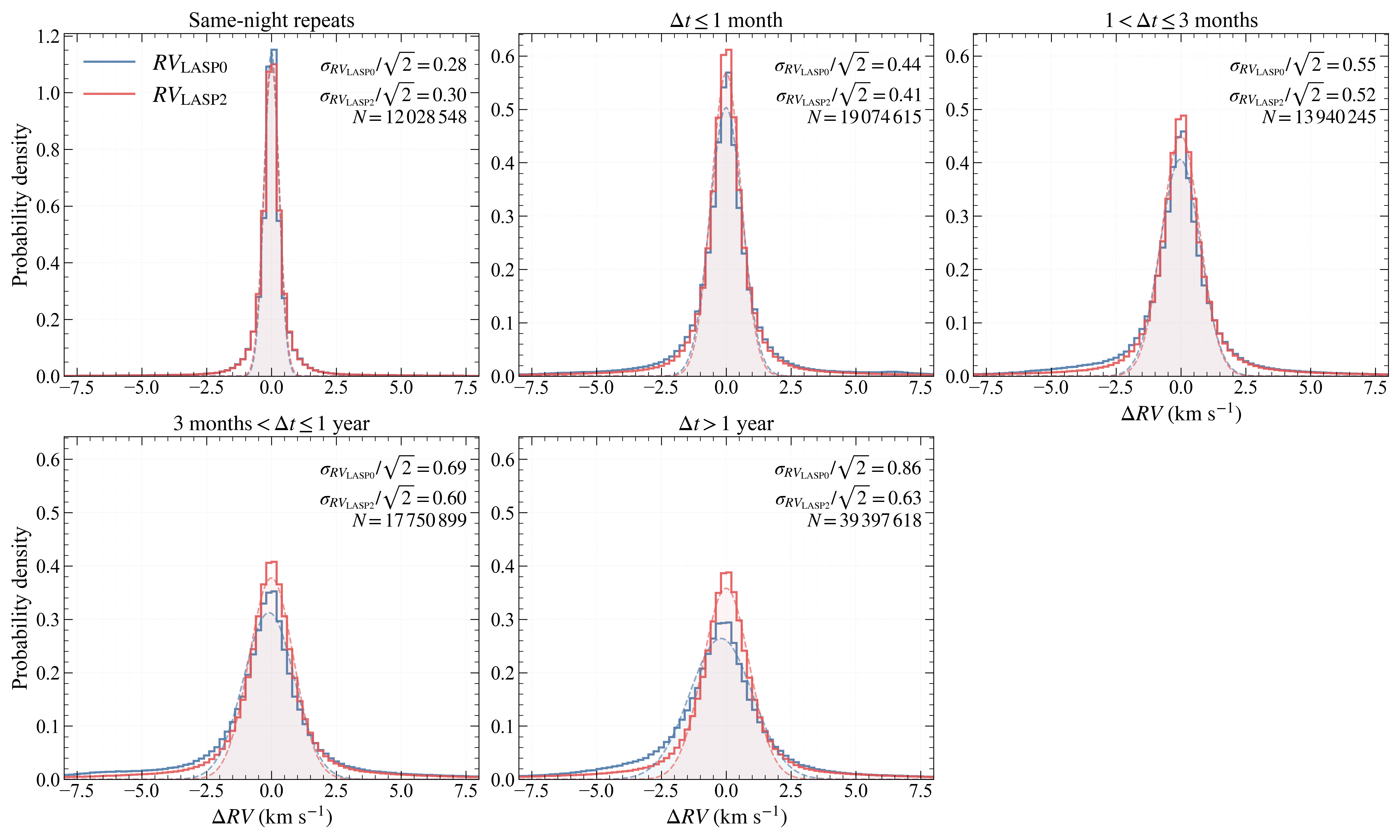}
\caption{Probability density of pairwise RV differences for repeat observations of the same source, stratified by time interval between visits. From left to right and top to bottom, the panels show same-night repeats, $\Delta t \leq 1$\,month, $1 < \Delta t \leq 3$\,months, $3\,\text{months} < \Delta t \leq 1$\,year, and $\Delta t > 1$\,year. Blue and red histograms show the pre-correction $RV_{\rm LASP0}$ and the two-level-corrected $RV_{\rm LASP2}$ pairwise differences, respectively, each normalized to unit area. Dashed curves are Gaussian fits to the core of each distribution, with the fitted dispersion $\sigma$ divided by $\sqrt{2}$ to represent the single-measurement precision. $N$ denotes the number of pairs in each panel. The pre-correction scatter increases monotonically with time interval, reflecting growing zero-point systematics, while the two-level correction removes a progressively larger fraction of the systematic contribution.}

\label{fig:figure_6}
\end{figure*}

Figure~\ref{fig:figure_6} highlights this timescale dependence by grouping pairwise RV differences by the visit-to-visit time interval and reporting the corresponding single-observation precision. For same-night repeats, the scatter changes little after correction ($\sim$0.28--0.30\,km\,s$^{-1}$), consistent with both visits sharing a common instrumental state. With increasing baseline, the pre-correction single-observation scatter rises monotonically to $\sim$0.86\,km\,s$^{-1}$ at $\Delta t > 1$\,year, whereas the corrected scatter reaches only $\sim$0.63\,km\,s$^{-1}$. The two-level correction therefore removes an increasing fraction of the zero-point contribution on longer timescales, consistent with the spectrograph--exposure and fiber--time structure captured by the calibration.

\subsection{Validation using independent blue- and red-arm RV measurements}

In addition to the LASP pipeline, LAMOST DR12 provides RVs from a template cross-correlation pipeline using 483 selected Kurucz synthetic spectra, measuring the blue- and red-arm velocities separately ($RV_{\rm b0}$ and $RV_{\rm r0}$). While LASP reports RVs only for spectra with S/N$\geq10$, the cross-correlation RVs extend down to S/N$\approx2$, spanning a much wider dynamic range. They therefore offer an independent check that the two-level RVZP correction can be ported to a different RV pipeline and remains effective in the low-S/N regime not covered by LASP.

\begin{figure*}[!t]
\centering
\includegraphics[width=0.99\textwidth]{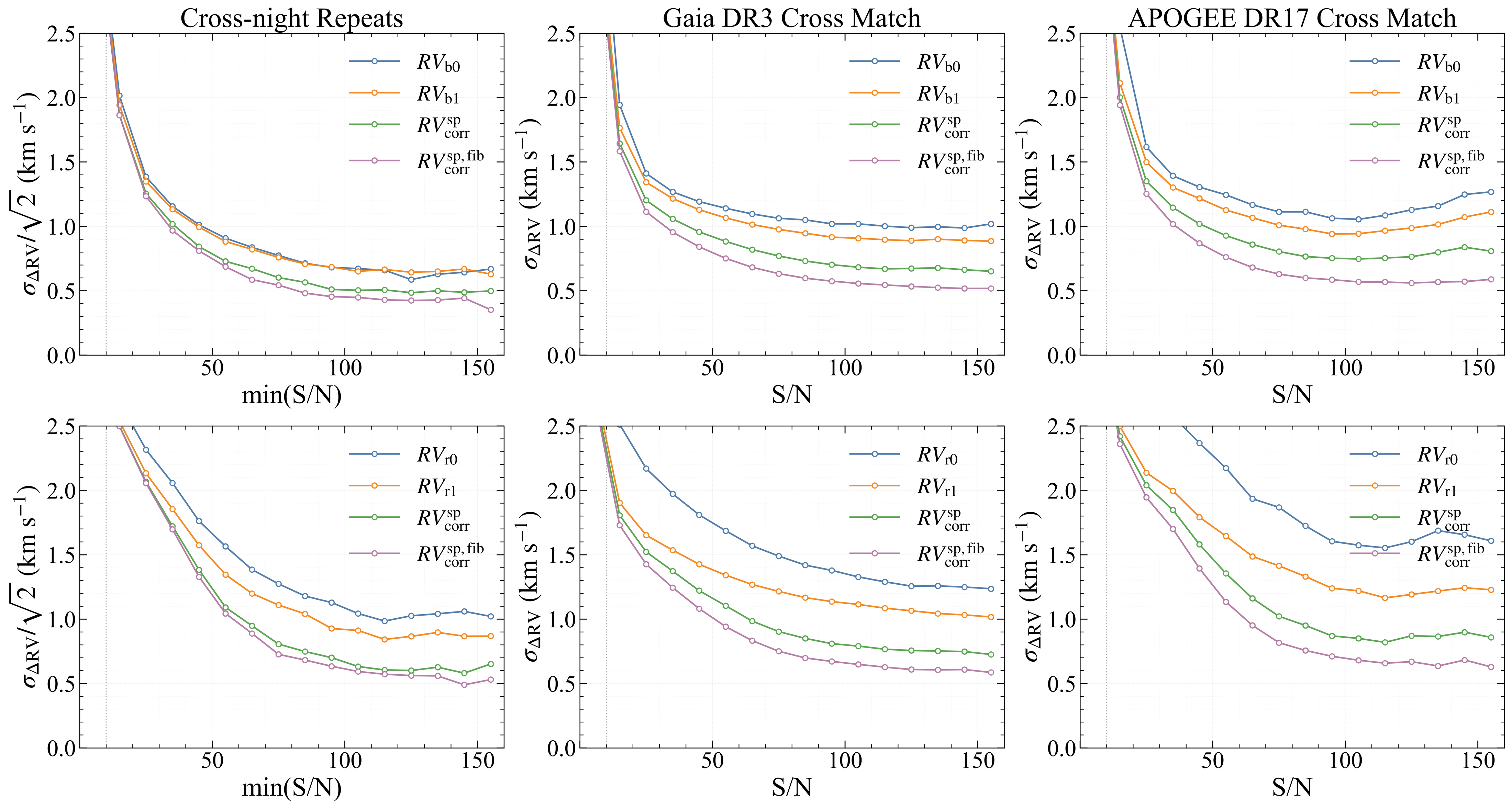}
\caption{Residual scatter versus LAMOST MRS S/N for blue- and red-arm RVs in three validation tests. Top/bottom rows show the blue/red arm; left to right: cross-night repeats, residuals relative to Gaia DR3 corrected RVs, and residuals relative to APOGEE DR17. Curves correspond to the raw RVs ($RV_{\rm b0}$ or $RV_{\rm r0}$; blue), the pipeline zero-point-corrected RVs ($RV_{\rm b1}$ or $RV_{\rm r1}$; orange), the spectrograph--exposure correction ($RV_{\rm corr}^{\rm sp}$; green), and the full two-level correction ($RV_{\rm corr}^{\rm sp,fib}$; purple). In each S/N bin, $\sigma$ is obtained from a Gaussian fit to the core of the residual distribution. For cross-night repeats, pairs are observations of the same source separated by $\geq365$\,days; observations from 2017 are excluded, and we require $|({\rm S/N})_1-({\rm S/N})_2|/\max[({\rm S/N})_1,({\rm S/N})_2]\leq0.30$. Pairs are binned by the lower S/N, and the pairwise scatter is divided by $\sqrt{2}$ to report the single-observation precision. For the Gaia and APOGEE comparisons, $\sigma$ is measured from LAMOST--reference RV differences and is not divided by $\sqrt{2}$; the Gaia sample is restricted to $\texttt{rv\_gaiadr3\_err}<0.5$\,km\,s$^{-1}$. The vertical dotted line marks S/N $=10$, the LASP lower limit. Results are shown up to S/N $=160$; the S/N $<10$ scatter lies above the plotted range and is reported in the text. The two-level correction improves both cross-night repeatability and agreement with the external references for both arms.}
\label{fig:arm_rv_validation}
\end{figure*}

We independently re-derive the full two-level correction for $RV_{\rm b0}$ and $RV_{\rm r0}$. All algorithmic choices are kept identical to those used for the LASP RVs. We assess the corrected velocities with three complementary tests: cross-night repeats, residuals with respect to the Gaia DR3 corrected RVs, and residuals with respect to APOGEE DR17. The repeat observations probe internal repeatability, while the two catalogue comparisons quantify external consistency. We extend the S/N binning to include S/N$<10$. The results are shown in Figure~\ref{fig:arm_rv_validation}.

All three comparisons show consistent improvements. At S/N$\geq10$, the impact of the two-level correction is qualitatively similar to that seen for the LASP RVs. At high S/N (S/N$=100$--160), the blue-arm scatter relative to APOGEE DR17 decreases from $\sim$1.14\,km\,s$^{-1}$ (raw) to $\sim$0.57\,km\,s$^{-1}$ (two-level corrected), compared with $\sim$1.00\,km\,s$^{-1}$ for the pipeline zero-point correction alone ($RV_{\rm b1}$). Over the same S/N range, the red-arm scatter decreases from $\sim$1.61 to $\sim$0.66\,km\,s$^{-1}$, compared with $\sim$1.22\,km\,s$^{-1}$ for $RV_{\rm r1}$. In nearly all S/N bins, the fully corrected RVs achieve smaller scatter (or better cross-night single-observation precision) than $RV_{\rm b1}$ and $RV_{\rm r1}$.

At S/N$<10$ (values above the plotted y-axis range in Figure~\ref{fig:arm_rv_validation}), the correction remains beneficial but the gain is smaller. For the blue arm, the residual scatter decreases from 4.37 to 3.58\,km\,s$^{-1}$ relative to Gaia and from 4.96 to 4.23\,km\,s$^{-1}$ relative to APOGEE DR17, while the cross-night single-observation precision improves from 3.78 to 3.63\,km\,s$^{-1}$. The corresponding red-arm improvements are 3.44 to 2.78\,km\,s$^{-1}$, 4.07 to 3.59\,km\,s$^{-1}$, and 3.52 to 3.40\,km\,s$^{-1}$. In this regime, the scatter is dominated by the random errors of the cross-correlation RVs. Since the instrumental zero-point term removed by our correction is only $\sim$0.5--1\,km\,s$^{-1}$ and adds in quadrature with measurement noise, only a modest reduction in total scatter is expected.

These results demonstrate that the two-level RVZP framework transfers directly to the blue- and red-arm cross-correlation RVs, without any modification. The correction remains helpful at low S/N, although the achievable precision there is limited primarily by measurement noise rather than residual instrumental zero-point systematics.

\subsection{Independent validation with APOGEE DR19}

\begin{figure}[htbp]
\centering
\includegraphics[width=\columnwidth]{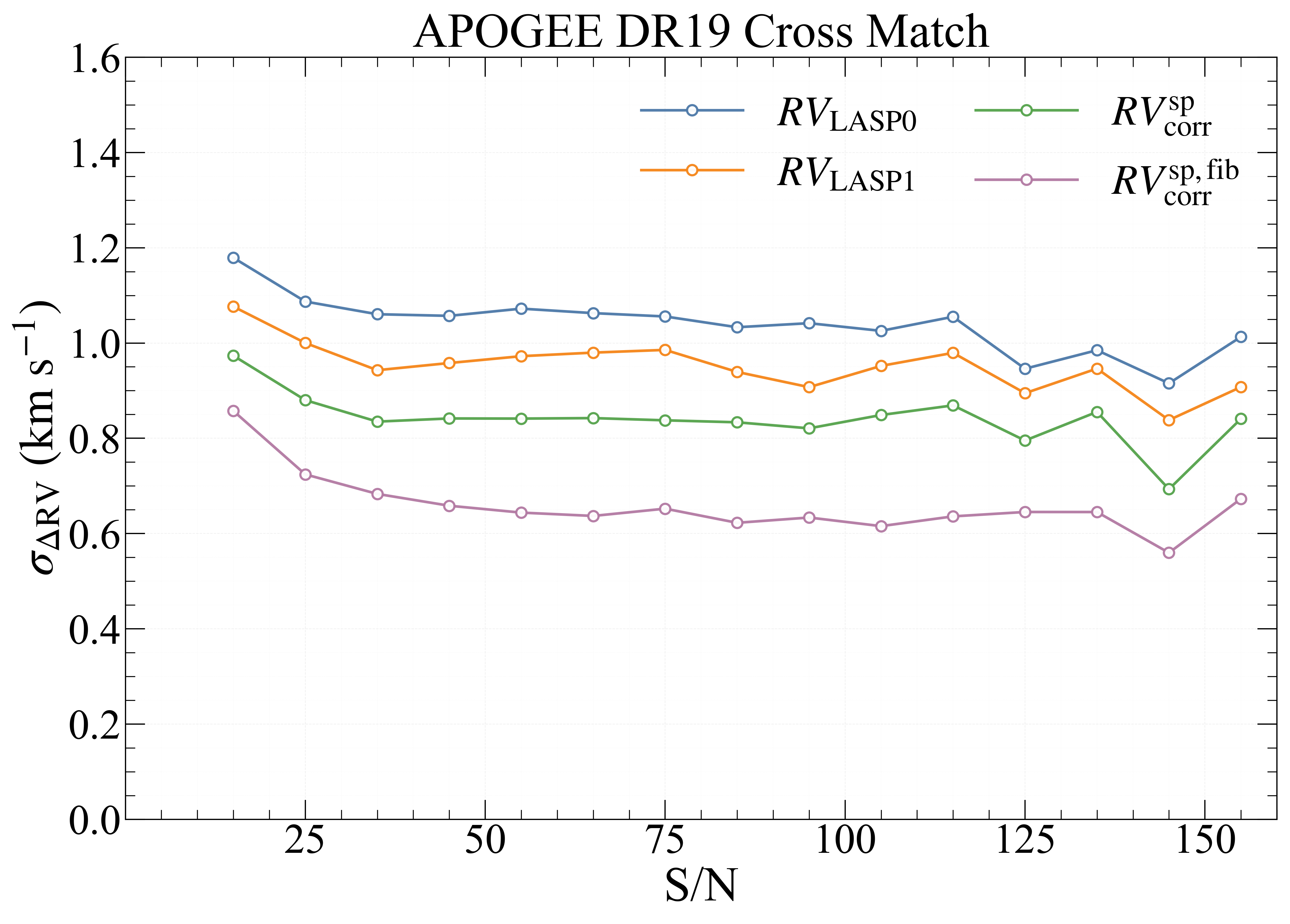}
\caption{Dispersion of $\Delta \mathrm{RV}$ versus S/N using APOGEE DR19 as an independent external validation. Because the DR19 sample is not used in the calibration, it provides an unbiased check of the achieved RV precision and the improvement from the two-level correction.}
\label{fig:figure_7}
\end{figure}

The APOGEE DR19 comparison sample is independent of our two-level correction: it is neither used in the calibration nor overlapping with the APOGEE DR17 set that defines the correction model. These RVs are drawn from the SDSS-V phase, which adopted the FPS infrastructure and an updated APOGEE reduction pipeline relative to SDSS-IV/APOGEE DR17 \citep{SDSSDR19}. With a large number of matches, DR19 provides a well-sampled external validation, so the reduced scatter represents a genuine out-of-sample test rather than a re-fit to the Gaia/APOGEE DR17 reference frame. Figure~\ref{fig:figure_7} compares $RV_{\rm LASP0}$, $RV_{\rm LASP1}$, $RV_{\rm corr}^{\rm sp}$, and $RV_{\rm corr}^{\rm sp,fib}$ to $RV_{\rm APOGEE\,DR19}$.

In the \texttt{S/N=50--60} bin, the scatter decreases from 1.07\,km\,s$^{-1}$ ($RV_{\rm LASP0}-RV_{\rm APOGEE\,DR19}$) and 0.97\,km\,s$^{-1}$ ($RV_{\rm LASP1}-RV_{\rm APOGEE\,DR19}$) to 0.84\,km\,s$^{-1}$ after the spectrograph--exposure correction and to 0.64\,km\,s$^{-1}$ after both correction levels, with the final product giving the lowest scatter. Note that the DR19 scatter is slightly larger than for DR17 (\ensuremath{\sim}0.64\,km\,s$^{-1}$ versus \ensuremath{\sim}0.52\,km\,s$^{-1}$ at high S/N).



\subsection{Uncertainty estimation}

The formal LASP RV uncertainties show an unphysical dependence on S/N: the median uncertainty is smallest at S/N$\sim$50 and then increases by $>$0.5\,km\,s$^{-1}$ toward higher S/N, instead of continuing to decrease. We therefore do not attempt to adjust the reported uncertainties; instead, we use the high-precision APOGEE DR17 RVs to calibrate empirical uncertainties for the corrected LAMOST RVs, treating the coadd$=0$ and coadd$=1$ samples separately.
In S/N bin $b$, we define the external residual as
\begin{equation}
\Delta RV_i = RV_{{\rm corr},i}-RV_{{\rm APOGEE\,DR17},i}.
\end{equation}
Following the procedure in Figure~\ref{fig:figure_5}, we iteratively apply $3\sigma$ clipping to the residuals in each S/N bin and fit the clipped distribution with a Gaussian. We adopt the fitted dispersion as the empirical RV uncertainty for that bin:
\begin{equation}
\sigma_{{\rm emp},b}=\sigma\left(\Delta RV_i\mid i\in b\right).
\end{equation}
For a spectrum with S/N in bin $b$, we assign the uncertainty of its corrected RV as
\begin{equation}
\sigma_{RV_{{\rm corr},i}}=\sigma_{{\rm emp},b(i)}.
\end{equation}

Because few sources have S/N$>180$, the dispersion in finer bins becomes unstable; we therefore group all spectra with S/N$>180$ into a single bin. 

The typical APOGEE DR17 RV uncertainty is $\approx0.1~{\rm km\,s^{-1}}$, much smaller than the dispersion of the LAMOST--APOGEE residuals. For simplicity, we do not subtract the APOGEE contribution from the residual variance; thus, $\sigma_{{\rm emp},b}$ should be interpreted as the typical external precision of normal stars in S/N bin $b$, not as a per-spectrum formal uncertainty. While this binned estimate does not capture small object-to-object differences, it directly reflects the observed scatter of the corrected LAMOST RVs relative to a high-precision external reference and provides a practical uncertainty scale for subsequent statistical analyses.

\section{Data Product}

The main data product is a value-added RV catalogue for the LAMOST DR12 v1.1 MRS blue-arm application sample. It contains 11,129,477 spectra in total (8,158,271 single-exposure and 2,971,206 coadded). The two-level corrected RV, \texttt{rv\_lasp2}, is provided for all entries with valid \texttt{rv\_lasp0}, including spectra that receive the correction but are not used to infer it.

For RV usage, \texttt{rv\_lasp0} is the LASP input RV, \texttt{rv\_lasp1} is the LASP static-corrected RV, and \texttt{rv\_lasp2} is the recommended RV for science analyses. The catalogue includes 47 columns in total; \texttt{coadd} distinguishes single-exposure from coadded spectra, and \texttt{flag} records calibration participation/usage status.

The catalogue is publicly available on Zenodo (\url{https://doi.org/10.5281/zenodo.21467895}).
The calibration pipeline used in this work, including the adopted $\sigma$-clipping thresholds and weighting parameters, is publicly available in a version-controlled GitHub repository (\url{https://github.com/CrotRest/lamost-mrs-rvzp}) and is permanently archived on Zenodo (\url{https://doi.org/10.5281/zenodo.21468183}).
Table~\ref{tab:catalog_columns} summarizes the column definitions.

\begin{table*}[htbp]
\centering
\renewcommand{\arraystretch}{0.85}
\begin{tabular}{clcl}
\hline\hline
Column & Name & Unit & Description \\
\hline
1  & \texttt{mobsid}              & --                    & Unique spectrum identifier (obsid + lmjm + band) \\
2  & \texttt{obsid}               & --                    & Unique observation identifier \\
3  & \texttt{uid}                 & --                    & Unique star identifier \\
4  & \texttt{gp\_id}              & --                    & Survey identifier of the source (Pan-STARRS, Gaia, or LAMOST) \\
5  & \texttt{designation}         & --                    & LAMOST target designation \\
6  & \texttt{obsdate}             & --                    & Observation date (yyyy-mm-dd) \\
7  & \texttt{lmjd}                & --                    & Local modified Julian day \\
8  & \texttt{mjd}                 & --                    & Modified Julian day \\
9  & \texttt{planid}              & --                    & Plan name \\
10 & \texttt{spid}                & --                    & Spectrograph ID \\
11 & \texttt{fiberid}             & --                    & Fiber ID \\
12 & \texttt{lmjm}                & --                    & Exposure identifier within a plan \\
13 & \textbf{\texttt{coadd}}      & \textbf{--}          & Coadd flag: 0 for single-exposure spectra and 1 for coadded spectra \\
14 & \texttt{ra\_obs}             & deg                   & Fiber pointing right ascension \\
15 & \texttt{dec\_obs}            & deg                   & Fiber pointing declination \\
16 & \texttt{snr}                 & --                    & Signal-to-noise ratio of the blue-arm spectrum \\
17 & \texttt{gaia\_source\_id}    & --                    & Gaia DR3 source identifier \\
18 & \texttt{gaia\_g\_mean\_mag}  & mag                   & Gaia $G$-band mean magnitude \\
19 & \texttt{gaia\_bp\_mean\_mag} & mag                   & Gaia $G_{\rm BP}$-band mean magnitude \\
20 & \texttt{gaia\_rp\_mean\_mag} & mag                   & Gaia $G_{\rm RP}$-band mean magnitude \\
21 & \texttt{fibertype}           & --                    & Fiber type (Obj or F-Std) \\
22 & \texttt{ra}                  & deg                   & Right ascension from input catalog \\
23 & \texttt{dec}                 & deg                   & Declination from input catalog \\
24 & \texttt{rv\_lasp0}           & km\,s$^{-1}$         & Heliocentric RV from the LASP (pre-correction) \\
25 & \texttt{rv\_lasp0\_err}      & km\,s$^{-1}$         & Uncertainty of \texttt{rv\_lasp0} \\
26 & \texttt{rv\_lasp1}           & km\,s$^{-1}$         & RV after static spectrograph correction by the LASP \\
27 & \texttt{rv\_lasp1\_err}      & km\,s$^{-1}$         & Uncertainty of \texttt{rv\_lasp1} \\
28 & \texttt{fibermask}           & --                    & Fiber mask flag \\
29 & \texttt{moon\_angle}         & deg                   & Angular distance to the Moon \\
30 & \texttt{lunardate}           & --                    & Lunar calendar date \\
31 & \texttt{moon\_flg}           & --                    & Moon contamination flag \\
32 & \texttt{teff\_lasp}          & K                     & Effective temperature from the LASP \\
33 & \texttt{teff\_lasp\_err}     & K                     & Uncertainty of effective temperature \\
34 & \texttt{logg\_lasp}          & dex                   & Surface gravity from the LASP \\
35 & \texttt{logg\_lasp\_err}     & dex                   & Uncertainty of surface gravity \\
36 & \texttt{feh\_lasp}           & dex                   & Metallicity from the LASP \\
37 & \texttt{feh\_lasp\_err}      & dex                   & Uncertainty of metallicity \\
38 & \texttt{vsini\_lasp}         & km\,s$^{-1}$         & Projected rotational velocity from the LASP \\
39 & \texttt{vsini\_lasp\_err}    & km\,s$^{-1}$         & Uncertainty of projected rotational velocity \\
40 & \texttt{alpha\_m\_lasp}      & dex                   & $[\alpha/\mathrm{M}]$ from the LASP \\
41 & \texttt{alpha\_m\_lasp\_err} & dex                   & Uncertainty of $[\alpha/\mathrm{M}]$ \\
42 & \texttt{objtype}             & --                    & Target type assigned by the observing plan \\
43 & \texttt{rv\_lasp2}           & km\,s$^{-1}$         & Two-level corrected RV (this work) \\
44 & \textbf{\texttt{rv\_lasp2\_err}}      & \textbf{km\,s$^{-1}$}         & Uncertainty of \texttt{rv\_lasp2} \\
45 & \texttt{flag}                & --                    & Calibration usage flag: \\
   &                              &                       & 0\,=\,used in both RVZP0 and RVZP1; \\
   &                              &                       & 1\,=\,used only in RVZP0; 2\,=\,used only in RVZP1; \\
   &                              &                       & 3\,=\,application-only (not used in calibration). \\
46 & \texttt{nref\_rvzp0} & -- & Number of reference stars in the spectrum's spectrograph-level (RVZP0) \\
   &                              &                       & calibration unit; \\
   &                              &                       & 0\,=\,no direct references; correction inherited from the grouped fallback. \\
47 & \texttt{nref\_rvzp1} & -- & As \texttt{nref\_rvzp0}, but for the fiber-level (RVZP1) unit \\
\hline
\end{tabular}
\caption{Column definitions for the value-added catalogue (11,129,477 spectra: 8,158,271 single-exposure and 2,971,206 coadded). The \texttt{flag} column records calibration usage, and \texttt{rv\_lasp2} is the recommended RV for science analyses.}
\label{tab:catalog_columns}
\end{table*}

\section{Conclusion}

We present a two-level RVZP correction framework for LAMOST MRS using Gaia magnitude--colour-corrected RVs and APOGEE DR17 RVs as external references. Relative to the SEU/exposure-level time-dependent correction of \citet{zhangBO2021MRS}, our approach additionally models fiber--time residuals beyond the spectrograph--exposure level.

We summarize the main results as follows.
\begin{enumerate}
\item \textbf{Method.} We derive two empirical corrections: (i) a spectrograph--exposure term in \texttt{(lmjm, planid, spid)} units and (ii) a fiber--time term in \texttt{(spid, fiberid, time\_tag)} units (Figure~\ref{fig:figure_1}). For each unit, we combine MRS and reference uncertainties through inverse-variance weighting, apply outlier clipping, and estimate RVZPs with a weighted-median estimator to reduce sensitivity to RV-variable sources and locally anomalous reference measurements. For sparse units, a grouped baseline (``zero-level'') correction stabilizes the solution.

\item \textbf{Hierarchical RVZP structure.} The correction maps show that the MRS RV zero point is not a single global offset. At the spectrograph--exposure level, RVZPs vary coherently with observing season, date, and spectrograph at the km\,s$^{-1}$ level (Figure~\ref{fig:figure_2}). After this first-level correction, structured residuals remain at the fiber--time level as a function of fiber position and observing-time segment (Figure~\ref{fig:figure_4}).

\item \textbf{Precision and timescale dependence.} In the representative \texttt{S/N = 50-60} bin, the cross-night single-observation precision improves from 0.60\,km\,s$^{-1}$ to 0.50\,km\,s$^{-1}$ after the spectrograph--exposure correction and to 0.48\,km\,s$^{-1}$ after the full two-level correction (Figure~\ref{fig:figure_5}). The improvement grows with visit-to-visit time separation: for \(\Delta t>1\) year, the single-observation precision improves from 0.86 to 0.63\,km\,s$^{-1}$ after the full correction (Figure~\ref{fig:figure_6}).
Beyond the improved internal repeatability, the agreement with external references improves even more (Figure~\ref{fig:figure_5}): the scatter relative to Gaia DR3 corrected RVs decreases from 1.07 to 0.60\,km\,s$^{-1}$, and that relative to APOGEE DR17 decreases from 1.09 to 0.52\,km\,s$^{-1}$ (a $\sim$2\,$\times$ improvement).

\item \textbf{Independent validation.} We validate the correction against APOGEE DR19, which is not used in the calibration. In the \texttt{S/N = 50--60} bin, the scatter decreases from 1.07\,km\,s$^{-1}$ for \(RV_{\rm LASP0}\) and 0.97\,km\,s$^{-1}$ for \(RV_{\rm LASP1}\) to 0.84\,km\,s$^{-1}$ after the spectrograph--exposure correction and to 0.64\,km\,s$^{-1}$ after the full correction, indicating that the solution generalizes beyond the calibration references (Figure~\ref{fig:figure_7}).

\item \textbf{Data product and applications.} The value-added catalogue provides two-level corrected RVs for 11,129,477 blue-arm spectra (8,158,271 single-exposure and 2,971,206 coadded). It supplies a uniform RV baseline for applications requiring stable multi-epoch velocities, including Galactic kinematics, star clusters, binaries \citep{guo2025sb2orbits,he2025algol,zhang2025detachedeb,guo2026contact11,li2026contact146}, and time-domain RV studies.
\end{enumerate}

In summary, this work extends the velocity-calibration framework of \citet{zhang2026lrs} to LAMOST MRS and introduces an explicit correction for fiber--time RV zero-point systematics. The full two-level correction delivers substantially improved internal and external RV consistency across exposures, spectrographs, fibers, and epochs.

\vspace{7mm} \noindent {\bf Acknowledgments}
{\bf The authors thank the anonymous referee for his/her suggestions that improved the quality of the paper.}
This work is supported by the National Key R\&D Program of China via 2024YFA1611901 and 2024YFA1611601, and the National Natural Science Foundation of China through the projects 12222301, and 12173007.

This work made use of the data from LAMOST (Large Sky Area Multi-Object Fiber Spectroscopic Telescope, also known as the Guoshoujing Telescope) (\url{https://cstr.cn/31118.02.LAMOST}). LAMOST is a Chinese national mega-science facility, operated by National Astronomical Observatories, Chinese Academy of Sciences.

This work has made use of data from the European Space Agency (ESA) mission {\it Gaia} (\url{https://www.cosmos.esa.int/gaia}), processed by the Gaia Data Processing and Analysis Consortium (DPAC, \url{https://www.cosmos.esa.int/web/gaia/dpac/consortium}). Funding for the DPAC has been provided by national institutions, in particular the institutions participating in the Gaia Multilateral Agreement.

\appendix
\setcounter{figure}{0}
\renewcommand{\thefigure}{A\arabic{figure}}

\section{Coadded-spectra RVZP calibration}

This appendix reports the RVZP structure and validation for the coadded MRS spectra. We apply the same two-level framework as for the single-exposure spectra, but derive an independent coadd-specific solution. Because coadds are constructed from the same underlying exposures, the qualitative RVZP patterns and validation trends are similar, although the numerical values can differ.

\begin{figure*}[htbp]
\centering
\includegraphics[width=0.95\textwidth]{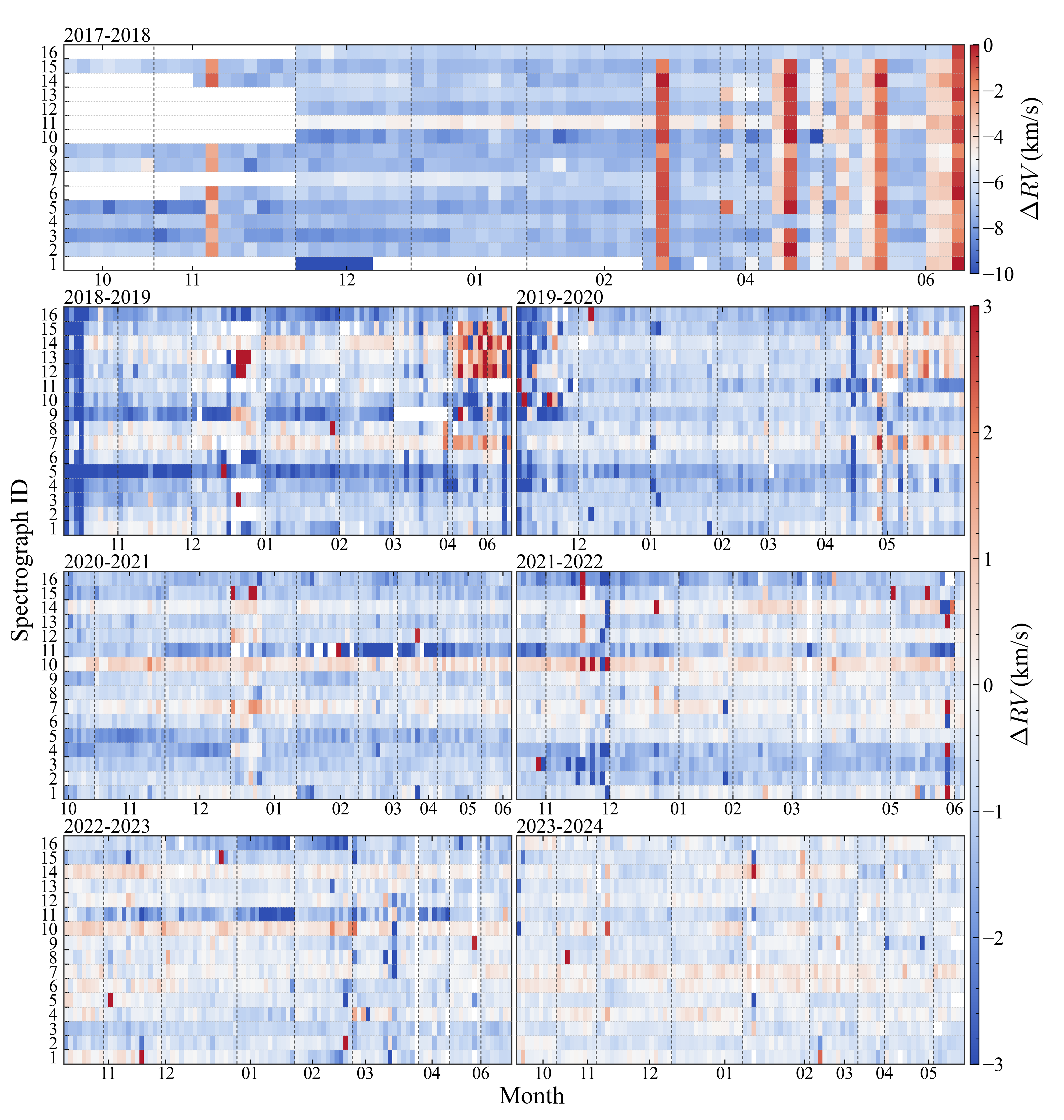}
\caption{Same as Figure~\ref{fig:figure_2}, but for the coadded spectra.}
\label{fig:appendix_a1}
\end{figure*}

\begin{figure*}[htbp]
\centering
\includegraphics[width=0.95\textwidth]{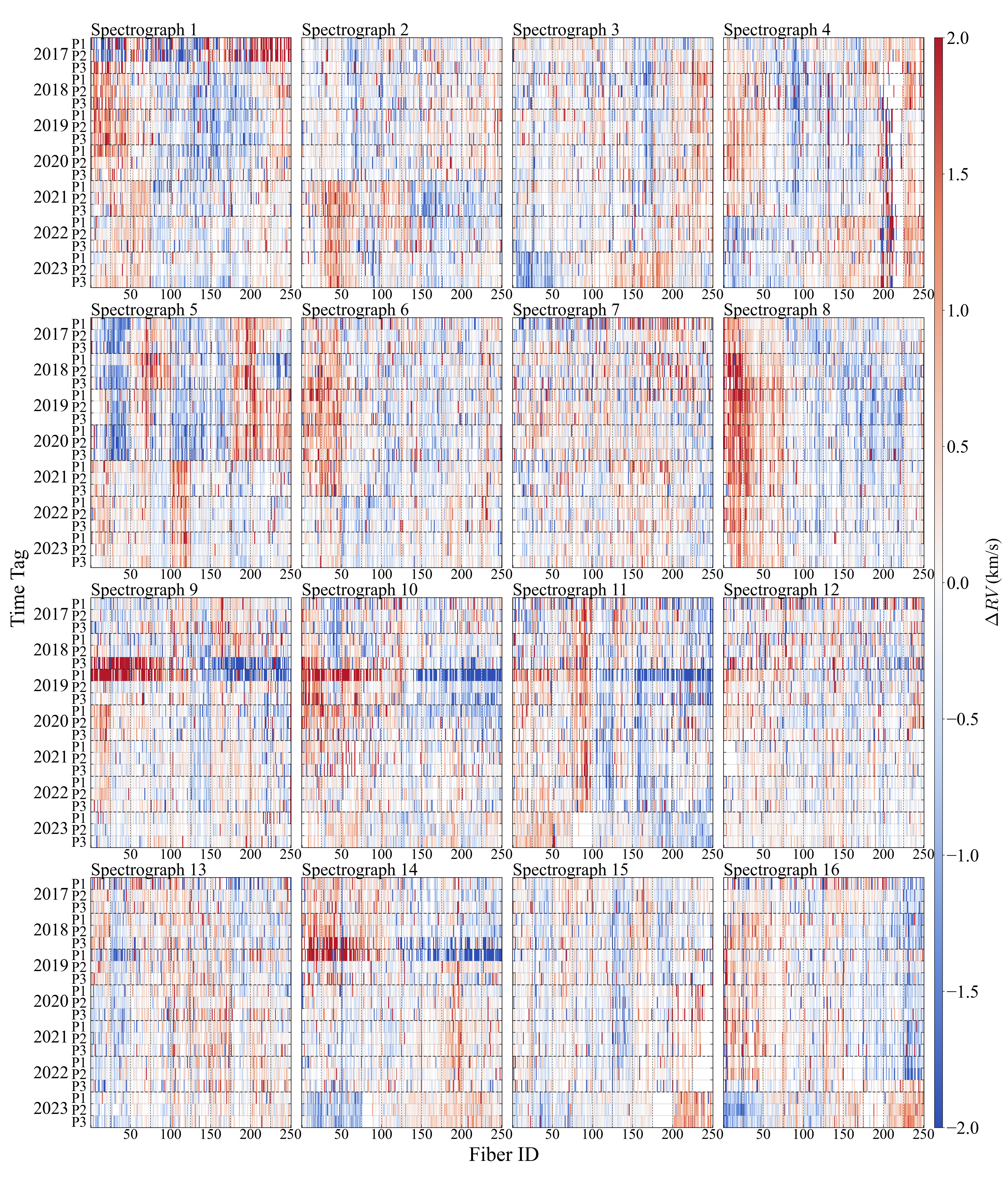}
\caption{Same as Figure~\ref{fig:figure_4}, but for the coadded spectra.}
\label{fig:appendix_a2}
\end{figure*}

Figure~\ref{fig:appendix_a1} shows the spectrograph--exposure RVZP structure for the coadds (cf. Figure~\ref{fig:figure_2}), demonstrating that the time- and spectrograph-dependent zero-point structure persists in the coadded product. Figure~\ref{fig:appendix_a2} shows the corresponding fiber--time RVZP map (cf. Figure~\ref{fig:figure_4}), indicating that residual spatial--temporal structure remains and motivates the second-level correction.

Comparing zero-point values derived independently from the coadded and single-exposure spectra, we measure median offsets of \(+0.006~\mathrm{km\,s^{-1}}\) and \(0.000~\mathrm{km\,s^{-1}}\), with dispersions of \(0.173\) and \(0.237~\mathrm{km\,s^{-1}}\) for the spectrograph- and fiber-level corrections, respectively, demonstrating that the two calibration solutions agree within the measurement scatter.

\begin{figure*}[htbp]
\centering
\includegraphics[width=0.95\textwidth]{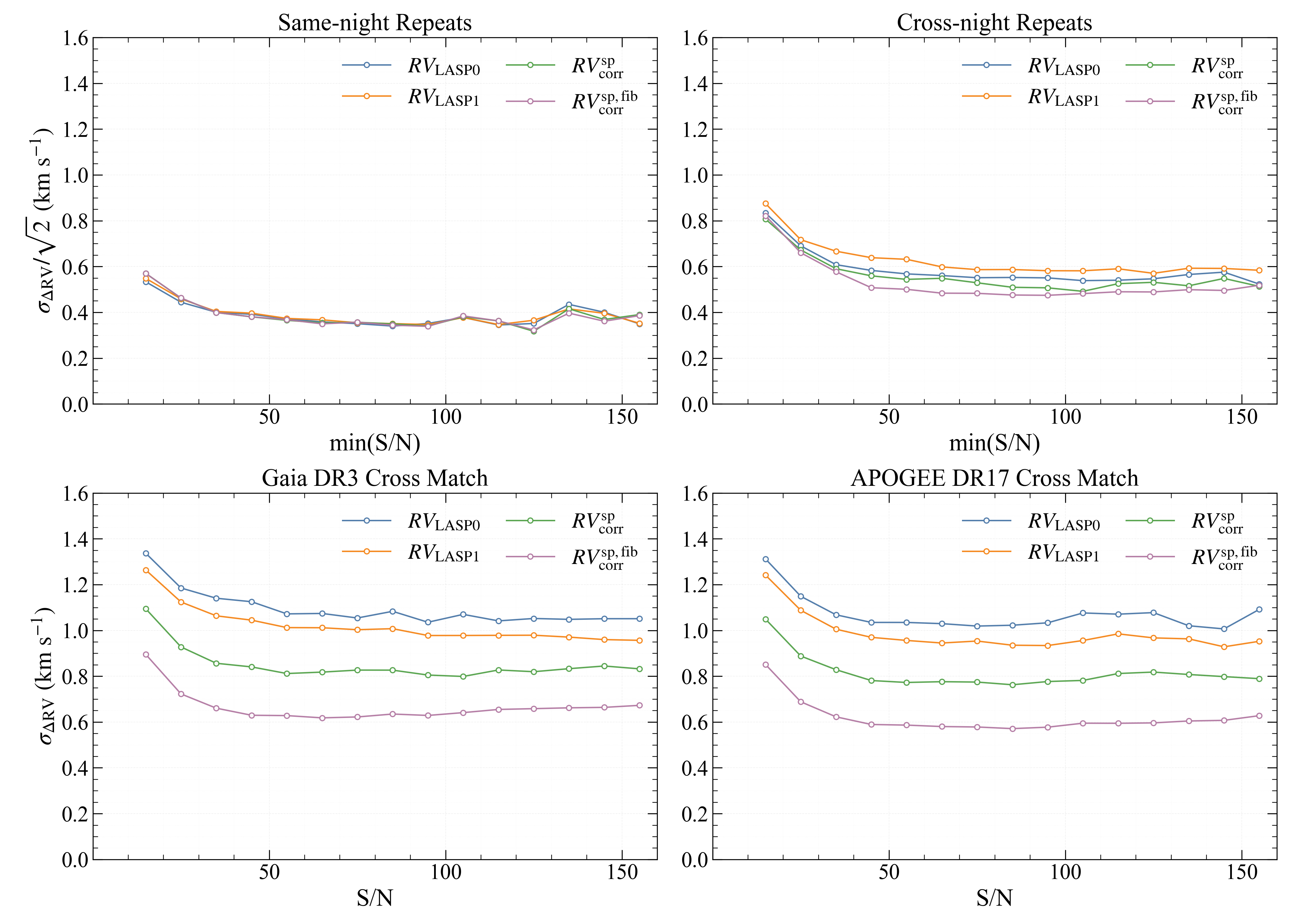}
\caption{Same as Figure~\ref{fig:figure_5}, but for the coadded spectra.}
\label{fig:appendix_a3}
\end{figure*}

\begin{figure}[htbp]
\centering
\includegraphics[width=0.5\columnwidth]{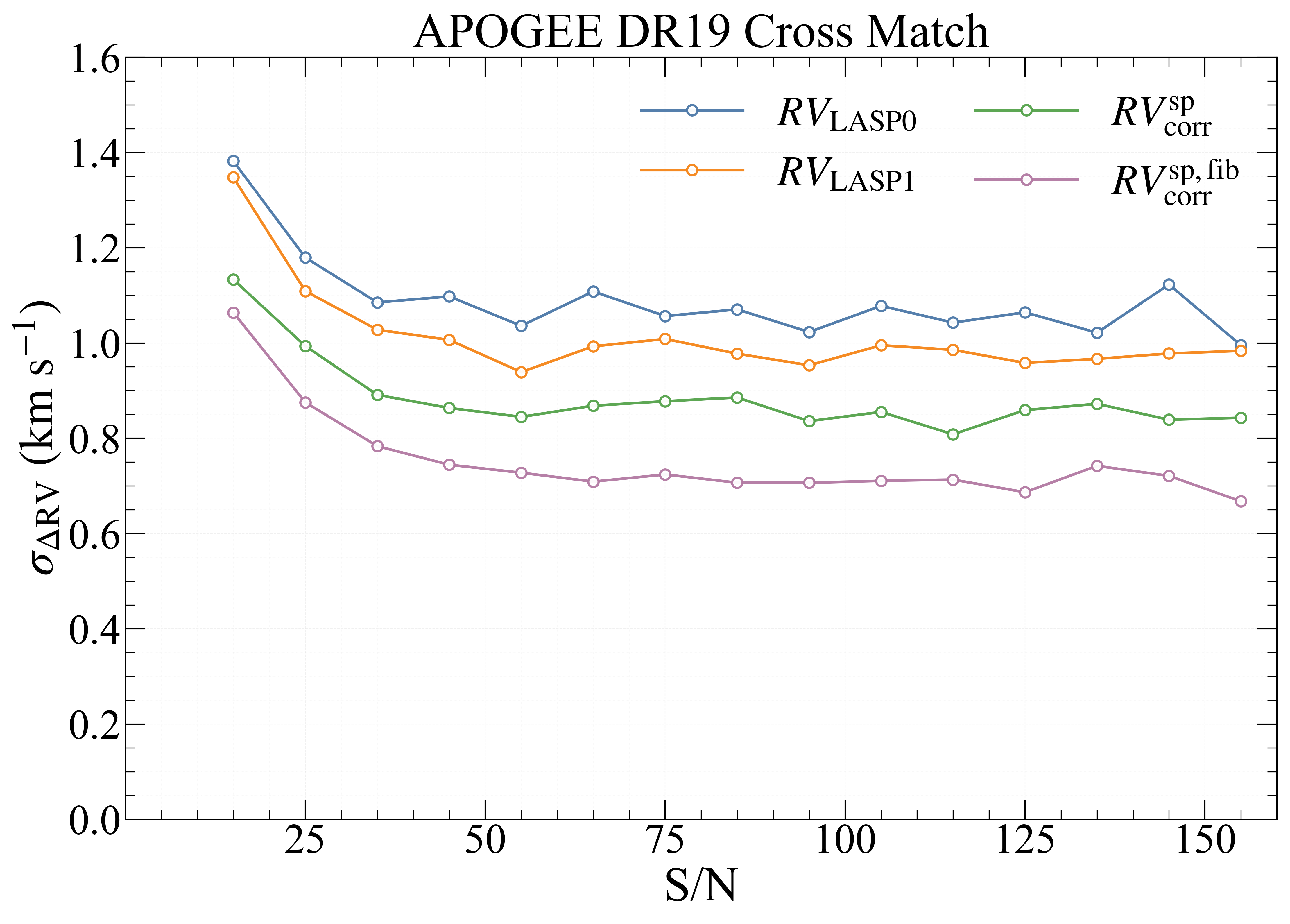}
\caption{Same as Figure~\ref{fig:figure_7}, but for the coadded spectra.}
\label{fig:appendix_a4}
\end{figure}

Figures~\ref{fig:appendix_a3} and \ref{fig:appendix_a4} present the S/N-dependent validation and the independent APOGEE DR19 check for the coadds (cf. Figures~\ref{fig:figure_5} and \ref{fig:figure_7}). In Figure~\ref{fig:appendix_a3}, the residual scatter after applying the second-level fiber correction is slightly larger for the coadded spectra than for the single-exposure spectra. This is expected because the coadd solution is derived on the same fiber--time grid but with roughly a factor of three fewer measurements per calibration unit, leading to a modest increase in calibration uncertainty. Overall, the coadd results reinforce the conclusions from the single-exposure spectra: the RV zero point is hierarchical, a two-level correction is required, and the corrected RVs show substantially improved internal and external consistency.

\bibliography{ref}
\bibliographystyle{aasjournal}

\end{document}